\documentclass[%
superscriptaddress,
reprint,
 bibnotes,
 amsmath,amssymb,
 aps,
longbibliography,
]{revtex4-1}

\usepackage{graphicx}% Include figure files
\usepackage{dcolumn}% Align table columns on decimal point
\usepackage{bm}% bold math
\usepackage[usenames,dvipsnames]{xcolor} % mk added

\usepackage[colorlinks,citecolor=MidnightBlue,linkcolor=Orange,urlcolor=MidnightBlue]{hyperref}% add hypertext capabilities

\usepackage[mathlines]{lineno}% Enable numbering of text and display math
\usepackage{color}
\usepackage{MnSymbol}
\definecolor{orange}{rgb}{1.0,0.4,0.0}
\definecolor{rose}{rgb}{1.0, 0.33, 0.64}
\usepackage{tikz}
\usepackage{times} % mk added
\usepackage{rotating}
\usepackage{gensymb}
\usepackage{siunitx}
\usepackage{xr-hyper}
\usepackage{commath}
\usepackage[colorlinks,citecolor=MidnightBlue,linkcolor=Orange,urlcolor=MidnightBlue]{hyperref}
\usepackage{cleveref}

\begin{document}

\newcommand{\blueline}{\raisebox{2pt}{\tikz{\draw[-,black!10!blue,solid,line width = 1pt](0,0) -- (3mm,0);}}}

\newcommand{\blackline}{\raisebox{2pt}{\tikz{\draw[-,black!40!black,solid,line width = 1pt](0,0) -- (3mm,0);}}}

\newcommand{\magentaline}{\raisebox{2pt}{\tikz{\draw[-,black!5!rose,solid,line width = 1pt](0,0) -- (3mm,0);}}}

\title{Dynamical-state switching controls active microrod navigation in porous media}
%\title{History-dependent chirality tuning controls active microrod transport in porous media}
%title{Chirality controls navigation of active microrods in porous media}

\author{Ana Cazorla}
%\thanks{These authors contributed equally to this work}
\affiliation{Universidad de Granada, Applied Physics Departament, Modeling Nature Unit of Excellence, Granada, E-18071, Spain.}%
\author{Mar\'ia L. Jim\'enez}
\affiliation{Universidad de Granada, Applied Physics Departament, Modeling Nature Unit of Excellence, Granada, E-18071, Spain.}%
\author{Ra\'ul A. Rica-Alarc\'on}
\affiliation{Universidad de Granada, Applied Physics Departament, Modeling Nature Unit of Excellence, Granada, E-18071, Spain.}%
\author{Carla Fern\'andez-Rico}
\affiliation{Department of Materials, ETH Zürich, 8093 Zurich, Switzerland.}%
\author{Laura Alvarez}%
\email{laura.alvarez-frances@u-bordeaux.fr}
\affiliation{Univ. Bordeaux, CNRS, CRPP, UMR 5031, F-33600 Pessac, France.}%
%%TC:ignore

\begin{abstract}

Adaptive search and travel strategies enable biological swimmers to explore complex environments, but remain difficult to encode in minimal synthetic microswimmers. Here, we unveil multiple dynamical and orientational states of rod-like Janus microswimmers actuated via electric fields: planar active, oriented chiral and persistent rods. We discover a hidden dynamical state after pre-orienting the rods, which allows us to switch on demand from strong chiral to persistent motion. We harness this orientation and chirality switch for exploration in microstructured porous arrays. Here, strong chirality promotes local scanning but drives circular trapping and poor net displacement, whereas the pre-aligned dynamical state enables efficient obstacle sorting, and trap escape. Comparing experiments with simulations of active Brownian particles in the very same experimentally-extracted obstacle geometries, we quantify navigation efficiency via first-passage time statistics, revealing how chirality tuning optimizes the trade-off between local exploration and long-range transport.

\end{abstract}
%%TC:endignore
%\pacs{Valid PACS appear here}
\maketitle

%\section*{Introduction}

Navigating through disordered and complex environments is a fundamental requirement for microorganisms. From sperm cells reaching an egg to immune cells patrolling tissues, and bacteria foraging for nutrients in heterogeneous soils or mucus \cite{bhattacharjee2019, kurzthaler2021, bechinger2016, krummel2016,dentz2022}, the ability to move effectively through disordered media is essential for survival. To succeed, these organisms employ adaptive strategies that alternate between persistent directed motion, reorientation events to escape entrapment, and chirality or tumbling to probe their local surroundings \cite{berg1972,modica2022,karimi2025,mihlan2022,Dehkharghani2023}. %In particular, bacterial transport in porous media is strongly influenced by the regulation of persistence and reorientation dynamics in response to geometric confinement \cite{bhattacharjee2019, dentz2022, colin2017}.

Designing motility at the microscale to mimic biological microswimmers is a current open challenge critical for realizing next-generation microscopic vehicles for cargo transport, in particular targeted drug delivery \cite{sitti2015, bozuyuk2024, willems2025,jung2024,oltolina2020,moncho2020,gravan2025}, or non-invasive microsurgeries \cite{sitti2015, nelson2010,gervasoni2024,zhang2025,oltolina2020,gravan2025}. In this context, synthetic active colloids are widely used as a platform for engineering bioinspired microswimmers aiming at these applications \cite{sitti2015, Katuri2017, Palagi2018, murali2024, karimi2025}. Active colloids self-propel under diverse mechanisms \cite{bechinger2016}; however, unlike their biological counterparts, they typically lack an intrinsic ability to adapt their swimming strategy to environmental constraints. Moreover, their navigation in porous or complex media is often hindered by interactions with solid interfaces, leading to wall accumulation \cite{kilic2011, boymelgreen2015,ramos2016} or trapping in obstacles and blind pores due to steric and hydrodynamic effects \cite{volpe2011,miloh2013,boymelgreen2022}. At the same time, particle–interface interactions can be leveraged to guide motion and bias transport by designing obstacle arrangements or tuning particle–surface interactions \cite{Simmchen2016,Wykes2017,vanBaalen2025obstacles}. 

\begin{figure*}
    \centering
    \includegraphics[width=0.9\linewidth]{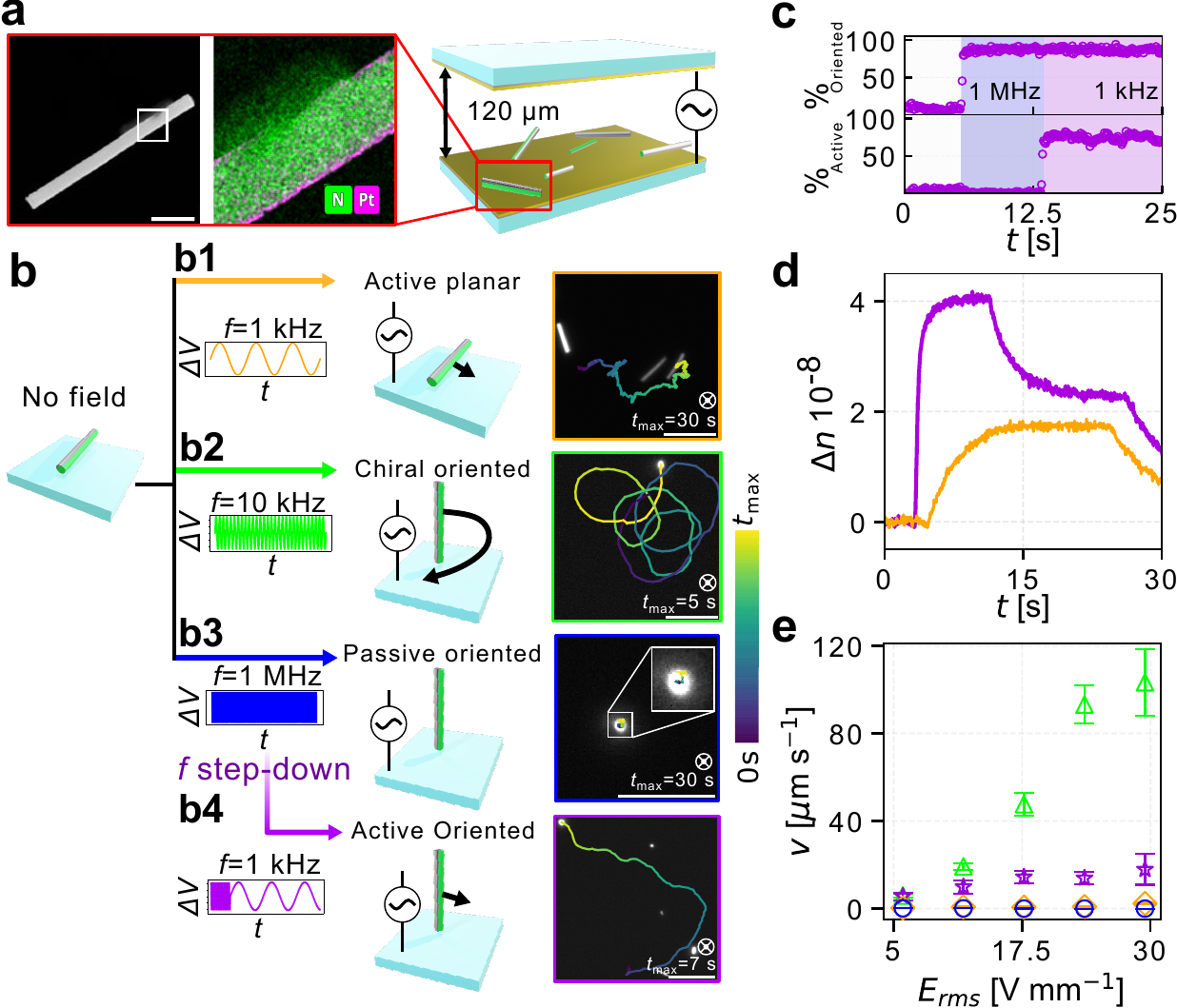}
    \caption{
\textbf{Frequency-dependent behaviour of Janus rods under AC electric fields.}
\textbf{a}  Plano-parallel electrodes set-up scheme. Inset: Scanning Transmission electron microscopy image of a Janus rod, and elemental analysis showing the spatial distribution of Nitrogen (green) and Platinum (magenta). Scale bar:  $\SI{2}{\mu m}$. 
\textbf{b} Diagrams and fluorescence microscopy images showing the behaviour under indicated AC electric field protocols (\textbf{b1}) $f=$\SI{1}{kHz} and  $\Delta V = \SI{10}{V_{pp}}$ ; (\textbf{b2}) $f=$\SI{10}{kHz} and  $\Delta V = \SI{10}{V_{pp}}$; (\textbf{b3}) $f=$\SI{1}{MHz} and  $\Delta V = \SI{8}{V_{pp}}$; and \textbf{b4} dropping $f$ from $f=\SI{1}{MHz}$ to \SI{1}{kHz} and  $\Delta V = \SI{10}{V_{pp}}$ (orientation locking protocol, OLP) (scale bars: \SI{10}{\mu m} in \textbf{b1, b2, b3} and \SI{20}{\mu m} in \textbf{b4}).
\textbf{c} Time evolution of oriented and active population of rod-like Janus colloids under OLP. 
\textbf{d} Birefringence signal ($\Delta n$) as a function of time. The orange curve represents the birrefringence for a $f = \SI{1}{kHz}$ pulse, while the purple curve shows it under the OLP application, where an initial high-frequency pulse ($f = \SI{1}{MHz}$) is brought to $f = \SI{1}{kHz}$ at $t \approx \SI{7.5}{s}$.
\textbf{e} Rod-like Janus colloids velocity $v$ as a function of the intensity of the applied field $E_{rms}$: orange squares for $f=\SI{1}{kHz}$; green triangles for $f=\SI{10}{kHz}$; blue circles for $f=\SI{1}{MHz}$; and for OLP (purple stars). Error bars are calculated as the standard error of the mean. 
}
    \label{fig:1}
\end{figure*}

Efficient navigation through porous and complex media, therefore, requires implementing refined motility strategies that include controlled reorientation mechanisms, which are indispensable for efficient dispersion in porous networks \cite{kurzthaler2021}. Inspired by biological navigation, current approaches to control dynamical patterns of active colloids use external fields to modulate propulsion and reorientation \cite{Colabrese2017,Karani2019,Sprenger2019,FernandezRodriguez2020}, or exploit internal material properties and geometry/composition reconfiguration to tune persistence length \cite{Ketzetzi2025} or mimic run-and-tumble behavior \cite{Karani2019,Alvarez2021,willems2025}.

On-demand control of chiral motion offers a minimal dynamical knob to reshape turning events, enabling a direct trade-off between local exploration and long-range persistence in synthetic microswimmers \cite{Alvarez2021}. In simulations of chiral active systems, obstacles can either enhance or hinder transport, with direct implications for controlling active motion in disordered media \cite{vanRoon2022}. Moreover, the interplay between angular velocity and alignment interactions can determine whether convex obstacles suppress or enhance surface accumulation \cite{vanRoon2025}. At larger (granular) scales, experiments with chiral active particles have shown that effective diffusivity becomes sensitive to the obstacle-lattice structure, a feature absent in achiral swimmers \cite{Chan2024}. At the microscale, however, chirality control is often not deliberate: fabrication imperfections can induce unintended symmetry breaking, generating persistent torques and non-adaptive or tunable chiral trajectories \cite{ebbens2010, lowen2016}. Although controlled chirality modulation has been achieved in bulk using feedback strategies and responsive materials \cite{Brooks2018, Alvarez2021, kesteren2023}, its use as an in situ control variable for navigation in porous media remains largely unexplored.

Here, we present a strategy that overcomes these limitations, enabling on-demand switching between chiral and persistent motion of colloidal rods using a frequency-switchable AC electric field. By identifying distinct frequency regimes, we successfully decouple electro-orientation from propulsion. This control suppresses the inherent chirality of the rods, converting circular trapping trajectories into persistent transport. We validate this approach in micro-structured porous media, demonstrating that the switching protocol allows rods to escape geometric traps and achieve optimal spreading, whereas constant-frequency ones lead to chiral swimmers remaining confined. This capability endows the colloids with a rudimentary form of physical intelligence \cite{sitti2021,gupta2021}, enabling a biomimetic search-and-travel mode fundamentally different from standard protocols that simply turn motility on or off.

\section*{Results}

\subsection*{Revealing hidden dynamical states via frequency pre-alignment}

In this work, we present the dynamical behavior and navigation efficiency of rod-like Janus active colloids under electric fields. We first fabricate the Janus rod particles by drying on a flat substrate a monolayer from a colloidal suspension of rod-like SU-8 particles \cite{fernandez2019} (diameter of $D=\SI{1}{\mu m}$, aspect ratio $ \langle L/D \rangle =4.6\pm2.0$). Then, we metalized one side with a platinum layer of $d_{\rm Pt}={6}$ nm using a sputter coater (see Methods for more details). After the sputter coating step, particles were immersed in MilliQ water and redispersed via ultrasonication. The asymmetry in particle coverage is confirmed by energy-dispersive X-ray spectroscopy imaging as shown in Fig.~\ref{fig:1}a. From the stock solution of the Janus rod-like particles, we take a volume and put it in a chamber of a height of  $z=\SI{120}{\mu m}$, in between two parallel electrodes as shown in Fig.~\ref{fig:1}a (see Methods for more details). Due to their size and density, the colloids sediment close to the bottom electrode, where they exhibit Brownian diffusion (see Supplementary Note 1, Fig.S1 and Supplementary Movie S1). Once all rod particles have sedimented, a perpendicular AC electric field is applied to investigate their dynamics at $f=1\,\mathrm{kHz}$, $10\,\mathrm{kHz}$, and $1\,\mathrm{MHz}$, and $E_{\rm rms}=2.95$--$\SI{29.47}{V\,mm^{-1}}$ ($\Delta V=$1--$\SI{10}{V_{\rm pp}}$); representative trajectories are shown in Fig.~\ref{fig:1}b1-b3.

For a fixed voltage difference of $\Delta V = \SI{10}{V_{pp}}$, we identify three dynamical states. At low frequency ($f\sim \SI{1}{kHz}$, Fig.~\ref{fig:1}b1 and Supplementary Movie S2) rods self-propel while lying parallel to the substrate in a planar orientation. As in their spherical Janus analogues, propulsion is dominated by asymmetric induced-charge electroosmotic (ICEO) flows around the particle \cite{Gangwal2008,Peng2014,FernandezRodriguez2020}, which generate electrohydrodynamic flows that drive the particle with the dielectric face leading \cite{Gangwal2008}. At intermediate frequency ($f\sim\SI{10}{kHz}$, Fig.~\ref{fig:1}b2 and Supplementary Movie S3), rods remain active but now electro-orient under a field-induced torque \cite{arenas2018}, and is also consistent with our previous work of rod-like colloids under the actuation of electric fields \cite{Fojo2025,cazorla2024}. The resulting field-aligned Janus rods move faster than the planar-oriented rods, due to reduced drag and substrate friction. Interestingly, they also exhibit strong chiral swimming behavior, most likely due to the anisotropy of the coating or the particle shape that magnifies when the particles stand up. At higher frequencies ($f \sim \SI{1}{MHz}$, Fig.~\ref{fig:1}b3 and Supplementary Movie S4), the rods retain their orientation along the applied AC field, but are no longer active due to the decrease of the ICEO flows driving the motion as $U\propto f^{-1}$ \cite{Yang2019}. 

When from this oriented and not motile state at $ f = \SI{1}{MHz}$, we abruptly decrease to $f = \SI{1}{kHz}$ (Fig.~\ref{fig:1}b4, $\Delta V = \SI{8}{V_{pp}}$), the particles retain their orientation and display active motion revealing a \textit{hidden dynamical state induced by pre-alignment} (Fig.~\ref{fig:1}c and Supplementary Movie S5). Hereafter, we refer to this procedure as the \textit{Orientation Locking Protocol} (OLP). In comparison with the oriented chiral state, they seem to be slower and also less chiral (as we will detail in the next section). This hidden state transition is characterized in Fig.~\ref{fig:1}c, where we quantify the evolution of the population of oriented and active particles. We observe an abrupt increase in the number of oriented particles at $f=\SI{1}{MHz}$, while no active particles are detected. This field was applied for a sufficient duration to ensure the particles were electro-oriented ($t\sim \SI{5}{s}$). When decreasing the frequency from $f=\SI{1}{MHz}$ (oriented not active) to $f=\SI{1}{kHz}$, the particles remain oriented for all the experimental time windows ($ t\gg \SI{60}{s}$), and most of the particles become active while oriented (Fig.~\ref{fig:1}c). We hypothesize that this behavior is driven by asymmetric electro-osmotic flows producing a restoring torque upon vertical tilting. This torque prevents the particles from diffusing and losing their alignment \cite{arenas2018}. 

To shed light on the underlying mechanism, we performed further measurements of the electro-orientation of the particles in the bulk experimental conditions under the mentioned OLP. The electric birefringence response (Fig.~\ref{fig:1}d) showed that after switching the field to 1 kHz, the birefringence $\Delta n$, which is proportional to the degree of alignment with the applied field (Supplementary Note S2, and Supplementary Fig.~S2), turns to nearly the same value as the direct protocol. This means that there is no effect of the pre-orientation of the birrefringence signal when comparing the OLP and direct frequency application with a final frequency of $f=\SI{1}{kHz}$. These results suggest that the asymmetry of the fluxes generated at the bottom of the chamber are necessary for the pre-orientation not to fade. 

\begin{figure}
    \centering
    \includegraphics[width=0.85\linewidth]{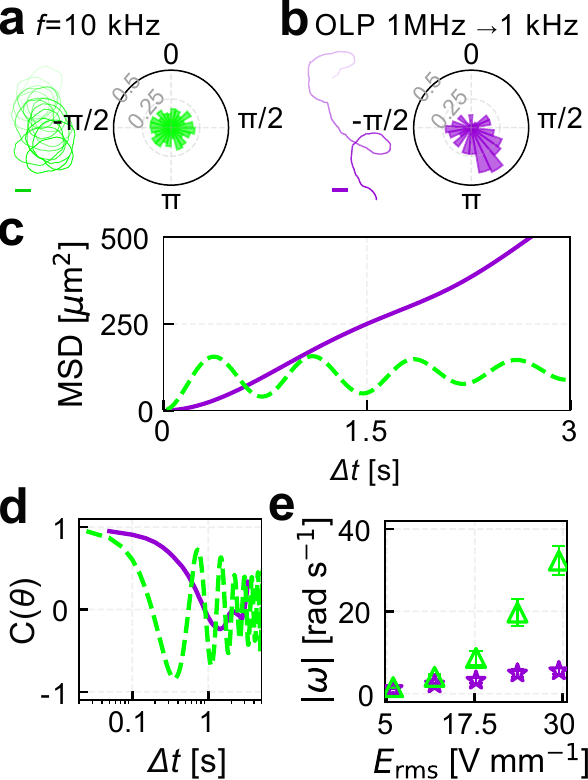}
    \caption{\textbf{Chirality swimming behavior under AC electric field.} 
\textbf{a, b} Representative trajectories (scale bars: \SI{5}{\mu m}) and corresponding rose diagrams of the displacement vector angle for rod-like Janus colloids under (\textbf{a}) $f = \SI{10}{kHz}$ field and (\textbf{b}) the OLP, both at $\Delta V = \SI{8}{V_{\text{pp}}}$.
\textbf{c–e} Comparison between $f = 10$ kHz AC field (green dashed line/triangles) and OLP (purple solid line/stars) for: (\textbf{c}) mean squared displacement (MSD), (\textbf{d}) autocorrelation of the displacement vector angle, and (\textbf{e}) angular velocity $\omega$ as a function of the applied field. Error bars represent the standard error of the mean.}
    \label{fig:3}
\end{figure}

\begin{figure*}
    \centering
    \includegraphics[width=0.9\linewidth]{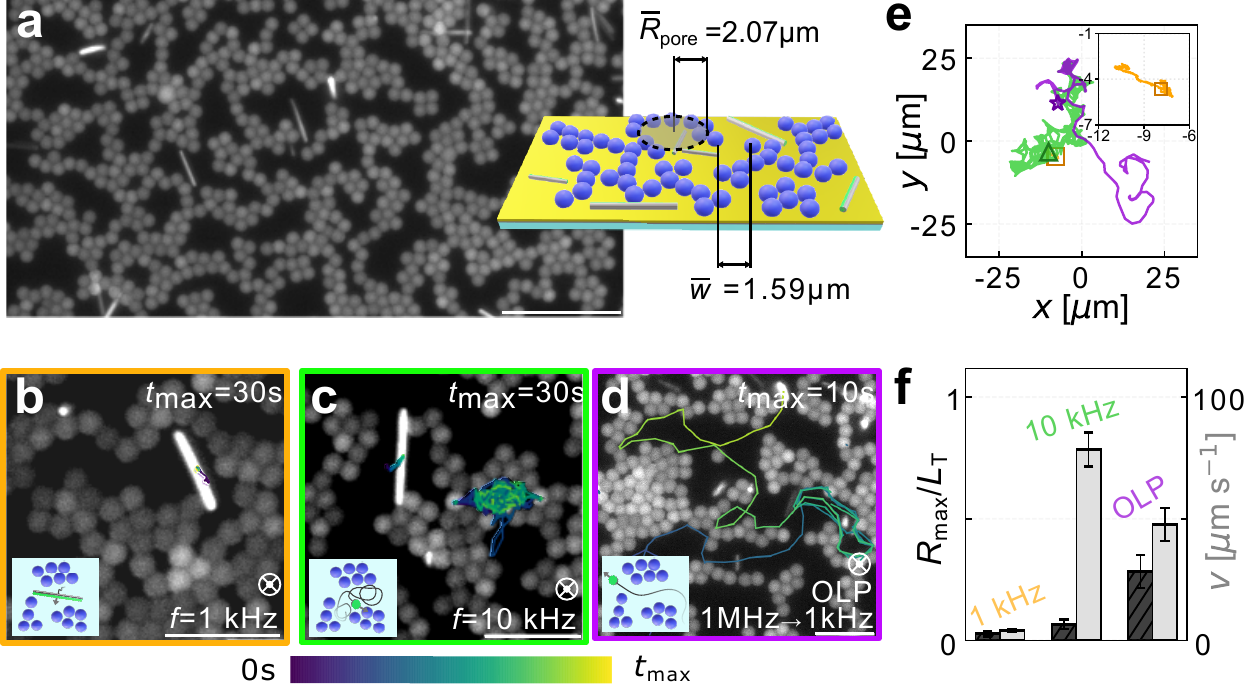}
    \caption{\textbf{Dynamics of rod-like Janus colloids in a porous media under AC electric fields.}  
\textbf{a} Fluoresce microscopy image of polystyrene microparticles built-up porous medium, scale bar $\SI{20}{\mu m}$. Inset: Schematic diagram of the 2D porous media via microparticle sedimentation on the bottom electrode of the chamber and pore sizes.
\textbf{b--d} Fluorescence microscopy image (scale bar $\SI{10}{\mu m}$), trajectory and diagram of the movement of a rod-like colloid in a porous medium while \textbf{(b)} applying a $f=\SI{1}{kHz}$ field, \textbf{(c)} a $f=\SI{10}{kHz}$ field and \textbf{(d)} OLP.
\textbf{e} Trajectory of a single particle undergoing sequently the following applied $\Delta V = \SI{8}{V_{pp}}$ AC electric field: with $f=\SI{1}{kHz}$ (yellow, starting point orange square), with $f=\SI{10}{kHz}$ (orange, starting point green triangle) and after performing OLP (purple, starting point purple star). Inset: zoom in of $f=\SI{1}{kHz}$ trajectory.
\textbf{f} Navigation performance for the three protocols as a function of the applied field. Left axis (dark gray): straightness index, defined as the maximum distance from the starting point normalized by the total path length. Right axis (light gray): Average instantaneous velocity. Error bars represent the standard error of the mean. 
}
    \label{fig:4}
\end{figure*}

The swimming speed $v$ for all of these behaviors at increasing voltage is depicted in Fig.~\ref{fig:1}e, and all of the active scales as $v\propto E^{2}$. Faster swimming rods are those in the chiral oriented state Fig.~\ref{fig:1}b2.
The active particles performing OLP from $f=\SI{1}{MHz}$ to $f=\SI{1}{kHz}$ exhibit longer persistence lengths $L_{\rm p}$, than their oriented analogues at $f=\SI{10}{kHz}$ (Supplementary Information, Supplementary Note S3 and Supplementary Fig.~S6). Therefore, after the abrupt frequency switching, the rods retain their orientation, which affects the system dynamics and reveals a hidden dynamical state. This allows for a fast change of dynamical patterns in space, which we will further explore later in the next section.

\subsection*{Chirality switching to navigate porous media}

We quantify in detail the dynamical behavior of the oriented-states of the active rod: i) directly actuated at $f=\SI{10}{kHz}$ (Fig.~\ref{fig:3}a), ii) and the oriented active Janus rod using OLP, changing from $f=\SI{1}{MHz}$ to $f=\SI{1}{kHz}$ (Fig.~\ref{fig:3}b), both at the same applied voltage. Upon direct application of $f=\SI{10}{kHz}$ the particle exhibits a chiral trajectory. However, the locking method induces a dynamical state with a persistent motion and a residual chirality. This difference in behavior is depicted in the rose plot of the angle distribution in Figs.~\ref{fig:3}a and b.

In Fig.~\ref{fig:3}c, we plot the mean squared displacement (MSD) for both types of dynamical conditions. When directly applying $f=\SI{10}{kHz}$, the MSD shows pronounced oscillations (characteristic of strong chiral motion) and approaches a plateau at long lag times, consistent with confinement to a finite orbit while the particle spins continuously. In contrast, under the OLP protocol, particles behave as an active Brownian particle (ABP) with weak chirality: the MSD exhibits a short-time ballistic regime \(\propto v^2 \tau^2\) followed by a crossover to an effective diffusive growth, with only a slight modulation at the chirality timescale. The strong oscillations under $f=\SI{10}{kHz}$ field also show up in the correlation function of the angle (Fig.~\ref{fig:3}d). 

We also measure the angular velocity, $\omega$, for various particles of each class ($N\approx 30$) in Fig.~\ref{fig:3}e. As with the linear velocity measurements, the angular velocity increases proportionally to the square of the field intensity for both cases and reach higher values for chirals colloids as expected. Furthermore, population statistics confirm that there is no preferential handedness for either protocol, resulting in a racemic mixture of trajectories (see Supplementary Note S3). Thus, with this abrupt change in frequency, we can keep the same orientational behavior, but in situ and on demand change the chirality behavior of the active Janus, which is particularly interesting for the exploration of porous media.

\begin{figure*}
    \centering
    \includegraphics[width=0.9\linewidth]{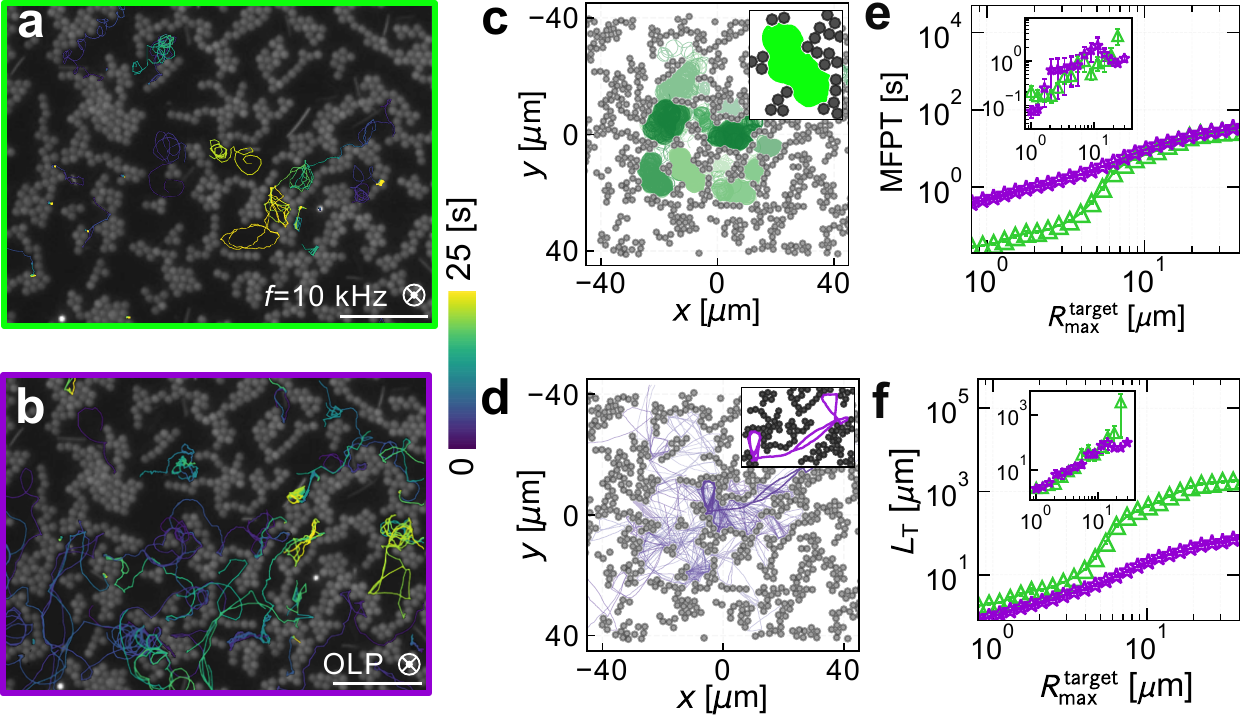}
    \caption{ \textbf{Experiments and simulation of active rods in porous media. }
\textbf{a, b} Fluorescence microscopy image of the porous medium (scale bar 10$\mu$m) and trajectories of rod-like Janus colloids \textbf{(a)} under $f=\SI{10}{kHz}$ field and \textbf{(b)} after applying a OLP. 
\textbf{c, d} Simulations of Active Brownian Particles (ABPs) in a reconstructed experimental porous medium (microparticle radius $R_{\rm mp} = \SI{1}{\mu m}$) with parameters matching (\textbf{c}) the $f = \SI{10}{kHz}$ and (\textbf{d}) the OLP, both at $\Delta V = \SI{8}{V_{pp}}$. The normalized effective displacement is $R_{\rm max}/L = 0.004$ and $R_{\rm max}/L_{\rm T} = 0.174$, respectively. Insets: Detail of a single simulated trajectory under each protocol.
\textbf{e, f} Comparison between $f = \SI{10}{kHz}$ field (green triangles) and OLP (purple stars) for 1000 simulated ABPs to reach a threshold distance ($R_{\rm max}^{\rm target}$): (\textbf{e}) Mean First Passage Time (MFPT) and (\textbf{f}) Path length ($L_{\rm T}$). Insets show the corresponding experimental data. Error bars represent the standard error of the mean, and solid lines are cubic spline interpolations. Error bars are calculated as the standard error of the mean. Solid lines are cubic spline interpolations.}
    \label{fig:5}
\end{figure*}

To quantify the transport performance of the different on-demand dynamical states in complex environments, we designed porous media using assembled microparticles. Briefly, a droplet of polystyrene spheres suspension with a radius of $R_{\rm mp}=\SI{1.0}{\mu m}$ was dried onto a slide, and the beads were subsequently sintered on the bottom electrode as depicted in Fig.~\ref{fig:4}a (more details about this protocol in Supplementary Note S4, Fig.S4, and Methods). Then, the parallel electrode configuration was assembled identically to previous experiments. When actuating the rods directly at $f=\SI{1}{kHz}$ (planar motion) they typically get stuck due to their bigger size with respect to the array characteristic constriction width (Fig.~\ref{fig:4}b and Supplementary Movie 6). When applying $f=\SI{10}{kHz}$ (oriented and highly chiral, Fig.~\ref{fig:4}c and Supplementary Movie 7), the particles effectively probe a smaller confined region than navigating the entire landscape,  despite their high swimming speed due to their higher chirality (with a minority of planar particles getting trapped). Finally, when applying the OLP 1 MHz–to–1 kHz (oriented and more persistent), the particles align efficiently with the field, easily escape from confined pores, and swim throughout the porous medium (Fig.~\ref{fig:4}d and Supplementary Movie 8). In Fig.~\ref{fig:4}e, we depict the trajectory of a single rod  within the porous array while changing the frequency and actuation sequence. While the particle remains hindered at $f=\SI{1}{kHz}$ and locally confined despite high speeds at $f=\SI{10}{kHz}$, the OLP enables effective obstacle sorting and particle movement.

In order to quantify their navigation efficiency, we calculated straightness index as the ratio of the longest distance from the starting point ($R_{\rm max}$) to the total path length traveled ($L_{\rm T}$), as well as the mean instantaneous velocity (Fig.~\ref{fig:4}f). Although the particles were faster under a $f=\SI{10}{kHz}$ field, those under the OLP stood out in the evaluation of the greatest $R_{\rm max}/L_{\rm T}$. Hence, we can clearly distinguish two regimes: one in which particles remain confined, probing small regions ($f=\SI{10}{kHz}$), and another in which particles can readily escape and travel between distant regions (OLP).

\subsection*{Mean-first passage time: a comparison to experiments}

\begin{figure}
    \centering
    \includegraphics[width=1\linewidth]{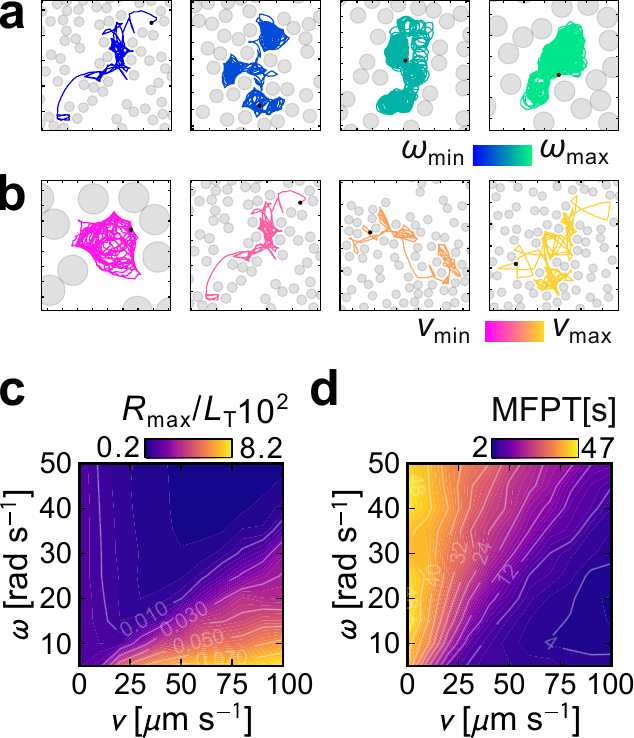}
    \caption{\textbf{Performance of simulated ABP in microparticle built-up porous arrays. }
\textbf{a} Trajectories of single ABP with different value of $\rm \omega$ and a fixed value of $v=\SI{40}{\mu m ~s^{-1}}$ for randomly generated porous arrays. 
\textbf{b}  Trajectories of single ABP with different value of $\rm v $ and a fixed value of $\omega = \SI{5}{rad ~s^{-1}}$ for randomly generated porous arrays.
\textbf{c} Heatmap of mean  $R_{\rm max}/L_{\rm T}$ as a function of the angular velocity $\omega$ and the linear velocity $v$  ($N_{\rm ABP}$=500).
\textbf{d} Heatmap of MFPT for $R_{\rm max}^{\rm target}=\SI{10}{\mu m}$ as a function of the angular velocity $\omega$ and the linear velocity $v$  ($ N_{\rm ABP}$=500). }
    \label{fig:6}
\end{figure}

To gain a comprehensive understanding of the role of chirality in exploring complex media, we performed numerical simulations to reproduce the experimental environment and extend the parameter space to conditions unattainable in our current system but relevant to future chiral particle design. In Fig.~\ref{fig:5}, we compare experimental observations (Fig.~\ref{fig:5}a and b) and numerical results obtained using the active Brownian particle (ABP) model~\cite{volpe2014} under two distinct conditions (Fig.~\ref{fig:5}c and d, see Methods for more details). The experimental porous environment was reconstructed by replicating the static obstacles directly from the experimental configurations. Steric interactions between the particles and the obstacle boundaries were explicitly accounted for via elastic collision, ensuring no overlap between the swimmers and the porous matrix. The first set of simulations reproduces the oriented chiral motion state at $f=\SI{10}{kHz}$ (Fig.~\ref{fig:5}c), using the experimentally measured values $ v = \SI{70}{\mu m ~s^{-1}}$ and $\omega = \SI{25}{rad ~s^{-1}}$ as input parameters. The second case mimics the 1 MHz–to–1 kHz OLP state (Fig.~\ref{fig:5}d), for which we adopt $v = \SI{20}{\mu m ~s^{-1}}$ and $\omega =\SI{0.5}{rad ~s^{-1}}$. This procedure was repeated for $N=1000$ independent trajectories, with the full particle paths recorded for post-processing (see Supplementary Note S5 for details).

Compared with the experiments, the simulated particles representing chiral motion state (Fig.~\ref{fig:5}c) appear less strongly trapped near the walls. This difference likely arises from the absence of short-range attractive interactions with the wall due to the AC field application. Nevertheless, in the simulations, the particles predominantly remain confined within a well-defined region over time due to their intrinsic chiral motion. In contrast, in the simulations corresponding to the OLP 1 MHz–to–1 kHz field state (Fig.~\ref{fig:5}d), particles escape most efficiently from the porous array, moving more persistently across the landscape.

To characterize the spatio-temporal exploration performance in complex environments, we report the Mean First Passage Time (MFPT) in Fig.~\ref{fig:5}e for simulations and experiments. This metric is defined as the average time required for a swimmer, starting from a random position, to first reach a predefined target displacement, $R_{\rm max}^{\rm target}$ \cite{khatami2016,yang2020}. Mathematically, for a single trajectory $\mathbf{r}(t)$, the first passage time is given by ${\tau}(R_{\rm max}^{\rm target}) = \min { (t > 0 : |\mathbf{r}(t)| \ge R_{\rm max}^{\rm target} )}$. Further details regarding these and other metrics used in this section can be found in Supplementary Fig.~S4a and Supplementary Note S5.

As observed in Fig.~\ref{fig:5}e, the high-chirality swimmer reaches the target distance ($R_{\rm max}^{\rm target} < \SI{6}{\mu m}=2v/\omega$) faster due to its larger propulsion speed. However, once the target distance exceeds the characteristic orbital diameter, the MFPT for the  $f = \SI{10}{kHz}$ regime exhibits a sharp upturn towards higher timescales comparable to the persistent swimmer. At these scales, the dynamical advantage of higher speed is canceled by the path tortuosity, which effectively traps the particle in a localized area. Consequently, at larger $R_{\rm max}^{\rm target}$, the dynamical effect of the higher speed is cancelled out, and both protocols converge to a similar long-time MFPT.

While the MFPT converges at large distances, the $L_{\rm T}$ reveals a fundamental difference between both strategies (Fig.~\ref{fig:5}f). To quantify this, we analyze the covered distance ($L_{\rm T})$ to reach $R_{\rm max}^{\rm target}$. As shown in Fig.~\ref{fig:5}f, the high-chirality swimmer consistently travels significantly larger $ L_{\rm{T}}$ to reach the same $R_{\rm max}^{\rm target}$. The continuous spinning induces frequent lateral deflections and wall encounters, forcing the particle to explore the void space redundantly (low straightness index, $R_{\rm max} / L_{\rm T} \ll 1$). In contrast, the persistent swimmer explores the porous landscape more efficiently through its openings despite its lower velocity. Thus, it achieves a ratio of displacement to path length closer to the high persistent motion limit ($R_{\mathrm{max}} / L_{\rm T}\to 1$). The separation in $L_{\rm{T}}$ is amplified, precisely when the MFPT values merge as chiral particles practically loop around a point. Although the simulations do not perfectly match all the experimental measurements (insets in \ref{fig:5} e and f), comparable MFPT values and divergence in $L_{\rm{T}}$ at large $R_{\rm{max}}^{\rm{target}}$ establish a robust framework. Consequently, we can exploit these simulations as a reliable tool to investigate further transport regimes beyond our current experimental reach. 

Using the numerical model, we further examine different random porous media and decouple the effect of swimming speeds and chirality (not possible in the experimental system) by extending the value of $v$ and $\omega$ to a broad range of values (Fig.~\ref{fig:6}). We simulate trajectories with a fixed linear $v$ and varying $\omega$ (Fig.~\ref{fig:6}a), and at fixed $\omega$ varying $v$ (Fig.~\ref{fig:6}b). We summarise our results of measuring the MFPT and $R_{\mathrm{max}} / L_{\rm T}\to 1$ for these two variables. In Fig.~\ref{fig:6}c, we identify two distinct transport regimes separated by a geometric threshold at $v/\omega \sim \SI{2}{\mu m}$. In the sub-pore regime ($v/\omega < \SI{2}{\mu m}$), particles undergo localized circular motion, and increasing linear velocity is counterproductive as it merely increases the redundant path length ($L_{\rm T}$) without facilitating escape. Therefore, the highest navigation efficiency is consistently obtained for the low-velocity particles. In the escape regime ($v/\omega > \SI{2}{\mu m}$), the trajectory radius exceeds the pore dimensions, forcing wall encounters that break the chiral confinement and enable long-range navigation. An overview of the navigation performance using the MFPT for $R_{\rm max}^{\rm target}=\SI{10}{\mu m}$ is provided in Fig.~\ref{fig:6}d. Consistent with the previous results, in the subporous regime ($v/\omega < \SI{2}{\mu m}$), high chirality increases exploration time by confining the particle regardless of its propulsion capacity. 

Our results suggest an interplay between the two magnitudes, revealing a crossover regime: particles with low linear and angular velocities can exhibit smaller MFPTs than those with higher values of both. Results for other values of $R_{\rm max}^{\rm target}$ can be found in Supplementary Fig.~S7 in Supplementary Note S5 .

\subsection*{Discussion}

The dynamical behavior and exploration strategy pattern of microswimmers is crucial for designing complex environment exploration \cite{volpe2017,sitti2015}. In particular, the ability to switch on-demand between persistent and chiral behavior provides a versatile mechanism to reconfigure microswimmer dynamics according to its specific goals and the local environmental constraints.

Here, we have presented a strategy to tune the orientation, swimming speed and chirality of Janus active rods actuated via AC electric fields only by switching the field frequency regime. Our work identifies and exploits the hidden dynamical state of the active rods changing from chiral to persistent motion while keeping a perpendicular orientation with respect to the plane. We therefore explore the navigation features of our system in experiments and simulations in porous media. We reveal a fundamental trade-off: while high chirality favors local exploration, it leads to entrapment. Conversely, the suppressed chirality achieved via the pre-alignment of the rod-like particles enables efficient long-range transport and obstacle avoidance.

This tunability, achieved here via frequency-switching AC electric field, could be a benchmark to design exploration pathways with other types of actuation mechanisms. The ability to switch in-situ between searching (chiral) and traveling (persistent) states mimics the diversity of swimming strategies seen in biological microswimmers, with precise external control in their synthetic analogues. This opens new avenues for applications in targeted cargo delivery within biological tissues, soil remediation, and the engineering of active materials that can self-regulate their transport properties.

\section*{Methods}
\subsection*{Fabrication of Janus rod-like particles}

To fabricate the rod-like active Janus particles, polymeric rods \cite{fernandez2019} were diluted in MilliQ water and deposited onto a glass slide, previously rendered hydrophilic by air plasma treatment. The suspension is sufficiently diluted to ensure a sub-monolayer where individual rods align with their long axes parallel to the substrate. Upon drying under controlled humidity, the monolayer is sputter-coated (Safematic CCU-010, Switzerland) with 6 nm of Pt. The particles are retrieved from the glass slide by sonicating the substrate in MilliQ water.

\subsection*{Cell preparation and AC electric field experiments}
The transparent electrodes are fabricated from 18mm $\times$ 18mm No.0 cover glasses ($85$-$115\mu$m -thick, Menzel Gl\"aser, Germany) covered with 3nm of Cr and 10nm of Au using a sputter coater (Safematic CCU-010, Switzerland). The bottom electrode is functionalized with a 10~mg/mL PSS (Sigma Aldrich, USA) solution to minimize particle adhesion to the substrate during the experiments. The solution is allowed to adsorb for 10 minutes before the electrode is rinsed and dried. A water droplet containing the particles is deposited on the bottom electrode within the $\SI{9}{mm}$ circular opening of a $\SI{0.12}{mm}$ thick sealing spacer (Grace Bio-Labs SecureSeal, USA). The top and bottom electrodes are connected to a signal generator (National Instruments Agilent 3352X, USA) that applies the AC electric field, varying frequencies from $f = \SI{1}{kHz}$ to $f = \SI{1}{MHz}$ and varying the voltage between $\Delta V=\SI{2}{V_{\rm pp}}$ and $\Delta V=\SI{10}{V_{\rm pp}}$.

%The transparent electrodes are fabricated from 18 mm $\times$ 18 mm No. 0 cover glasses ($85-115\,\mu $m-thick, Menzel Gl\"aser, Germany) covered with 3 nm of Cr and 10 nm of Au, deposited by metal evaporation (Evatec BAK501 LL, Switzerland), followed by 10 nm of SiO$_2$, deposited by plasma enhanced chemical vapour deposition (STS Multiplex CVD, UK) to minimize the adhesion of particles to the substrate during the experiments. A water droplet containing the particles is deposited on the bottom electrode within the 9 mm-circular opening of a 0.12 mm-thick sealing spacer (Grace Bio-Labs SecureSeal, USA). The top and bottom electrodes are connected to a signal generator (National Instruments Agilent 3352X, USA) that applies the AC electric field,  varying frequencies from 0.8 to 5 kHz and varying the voltage between 1 and 10 $V_{\rm PP}$.

\subsection*{Characterization of ABP dynamics}

%The Janus particles are imaged with a Zeiss Axio microscope in transmission and image sequences are taken with a sCMOS camera (Andor Zyla 4.2) at varying frame rates between 10 and 100 fps, depending on the swimming speed and concentration of the Janus particles, with a variable field of view among experiments. The positions of the centre of the JPs and the metal cap are located using Matlab routines. Then, a vector connecting both centres is used to determine the orientation of the particle at each frame. In addition, the centre of mass of the particle is used to determine the velocity vector. The image series of the metallic Janus particles actuated by the AC electric fields are acquired with a 20$\times$ and 40$\times$ objective (Zeiss). 

The Janus particles are imaged in bright field and fluorescence using an inverted microscope Olympus IX71 equipped with  $40\times$ objective. Image sequences are acquired with a sCMOS camera (Andor Zyla 4.2) at frame rates varying between 10 and 100fps, depending on the swimming speed and concentration of the particles. Particle trajectories were extracted using the TrackMate plugin in FIJI \cite{tinevez2017trackmate}. Subsequently, we measured the velocities of rod-like Janus particles by studying the mean squared displacement for small values of $\Delta t$ 
\begin{equation}
    {\text{MSD}}(\Delta t)=4D_{\rm t} \Delta t + v^2  \Delta t^2
\end{equation}
being $D_{\rm t}$ the translational diffusion coefficient and $v$ the propulsion velocity of the active colloid. Then, the persistence length ($L_{\rm p}$) which represents the distance it takes for an active colloid  to be statistically considered diffusive, is computed by fitting the ${\text{MSD}} (\Delta t)$ for greater values of $\rm \Delta t$:
\begin{equation}
    {\text{MSD}}(\Delta t)=4D_{\rm t} \Delta t + L_{\rm p}^2\left( D_{\rm r}\Delta t - 1 + \exp{(- D_{\rm r} \Delta t)} \right)
\end{equation}
being $D_{\rm r}$ the rational diffusion coefficient \cite{Bailey2022}.% The obtained values are represented in \ref{fig:1}c as a function of the intensity of the applied field . 

\subsection*{ABP Particle-based model}

The motion of the ABP is modeled using the overdamped Langevin equations \cite{volpe2014}. The particle's position vector $\vec{r}(t) = (x(t), y(t))$ and orientation angle $\theta(t)$ evolve over time according to:
\begin{equation}
	\dot{\vec{r}}(t) = v \hat{u}(\theta) + \sqrt{2D_{\rm t}} \vec{\xi}(t)
\end{equation}
\begin{equation}
	 \dot{\theta}(t) = \omega + \sqrt{2D_{\rm r}} \eta(t) 
\end{equation}
Where $v$ is the constant self-propulsion speed and $  \hat{u}(\theta) = (\cos \theta, \sin \theta)$ is the unit vector representing the orientation of the particle's propulsion axis. The term $\omega$ is a deterministic angular velocity, causing the particle to execute a persistent circular motion.

The terms $ \vec{\xi}(t)$ and $\eta(t)$ represent uncorrelated Gaussian white noise, accounting for thermal fluctuations. These noise terms have zero average and satisfy $ \langle \xi_i(t) \xi_j(t') \rangle = \delta_{ij} \delta(t-t')$ and $ \langle \eta(t) \eta(t') \rangle = \delta(t-t')$. The strengths of these fluctuations are determined by$D_{\rm t}$ and $D_{\rm r}$.

\section*{Data availability statement}
The data that support the findings of this study are available from the corresponding author upon reasonable request.

\section*{Code availability statement}
The code used in this study is available from the corresponding author upon reasonable request.

%\bibliography{reference}

\bibliography{reference_url}

\section*{Acknowledgements}

We thank José Fojo for the help with the first AC field experiment setup. L.A. thanks the French Agence Nationale de la Recherche (ANR-23-CE06-0007-0) and Frontiers of Life (U. Bordeaux) for funding. M.L.J.O. and R.R. thank Grant No. PID2024.161166NB.I00 funded by MICIU/AEI/10.13039/501100011033/FEDER/UE for funding. A.C. thanks Grant No. PRE2022-2-101943 funded by  MICIU/AEI /10.13039/501100011033 and by FSE+ for the financial support. 
%\textcolor{red}{SALVO QUE SEA IMPRESCINDIBLE, YO CREO QUE ES MÁS FÁCIL NO INDICAR QUIÉN AGRADECE A QUÉ PROYECTO, Y SOLO HACERLO PARA FUNDING INDIVIDUAL (BECAS), PERO COMO CONSIDERÉIS.}
\vspace{-0.5cm}

\section*{Author Contributions}
Author contributions are defined based on the CRediT (Contributor Roles Taxonomy) and listed alphabetically. Conceptualization: L.A.;  Formal analysis: A.C.,  Funding acquisition: L.A., M.L.J.O., R.R.; Investigation: L.A., A.C.; Methodology: L.A., A.C., C.F.R; Project administration: L.A.; Software: L.A., A.C.; Supervision: L.A., M.L.J.O., R.R.; Validation: L.A., M.L.J.O., R.R.; Visualization: L.A., A.C.; Writing – original draft: L.A., A.C. Writing – review, and editing: L.A., A.C. M.L.J.O., R.R., C.F.R.

\section*{Competing Interests}
The authors declare no competing interests.

\section*{Additional Information}
Supplementary Information is available for this paper. Correspondence and requests for materials should be addressed to L.A.

\end{document}